\documentclass[prx,nobalancelastpage,twocolumn,superscriptaddress,nofootinbib]{revtex4}

\usepackage{graphicx}
\usepackage{amssymb}
\usepackage{amsmath}
\usepackage[english]{babel}
\usepackage{color}
\usepackage[version=4]{mhchem}

\newcommand{\PMA}{Division of Physics, Mathematics and Astronomy, California Institute of Technology, Pasadena, CA 91125, USA}

\newcommand{\AP}{Thomas J. Watson, Sr., Laboratory of Applied Physics, California Institute of Technology, Pasadena, CA 91125, USA}
\newcommand{\AQT}{Alliance for Quantum Technologies, California Institute of Technology, Pasadena, CA 91125, USA}

\newcommand{\ket}[1]{ | #1 \rangle}

\begin{document}
\title{Telecom-band quantum optics with ytterbium atoms and silicon nanophotonics}
\author{Jacob P. Covey}\email{covey@caltech.edu}
\affiliation{\PMA}
%\affiliation{\IQIM}
\author{Alp Sipahigil}
%\affiliation{\IQIM}
\affiliation{\AP}
\author{Szilard Szoke}
%\affiliation{\IQIM}
\affiliation{\AP}
\author{Neil Sinclair}
\affiliation{\PMA}
%\affiliation{\IQIM}
\affiliation{\AQT}
\author{Manuel Endres}
\affiliation{\PMA}
%\affiliation{\IQIM}
\author{Oskar Painter}
%\affiliation{\IQIM}
\affiliation{\AP}

\begin{abstract}
Wavelengths in the telecommunication window ($\sim1.25-1.65$ $\mu$m) are ideal for quantum communication due to low transmission loss in fiber networks. To realize quantum networks operating at these wavelengths, long-lived quantum memories that couple to telecom-band photons with high efficiency need to be developed. We propose coupling neutral ytterbium atoms, which have a strong telecom-wavelength transition, to a silicon photonic crystal cavity. Specifically, we consider the $^3\text{P}_0\leftrightarrow^3\text{D}_1$ transition in neutral $^{171}$Yb to interface its long-lived nuclear spin in the metastable $^3$P$_0$ `clock' state with a telecom-band photon at $1.4$ $\mu$m. We show that Yb atoms can be trapped using a short wavelength ($\approx470$ nm) tweezer at a distance of 350 nm from the silicon photonic crystal cavity. At this distance, due to the slowly decaying evanescent cavity field at a longer wavelength, we obtain a single-photon Rabi frequency of $g/2\pi\approx100$\,MHz and a cooperativity of $C\approx47$ while maintaining a high photon collection efficiency into a single mode fiber. The combination of high system efficiency, telecom-band operation, and long coherence times makes this platform well suited for quantum optics on a silicon chip and long-distance quantum communication. 
\end{abstract}

\maketitle

\section{Introduction}
Efficient interfaces between single atoms and single photons could enable long-distance quantum communication based on quantum repeaters~\cite{Duan2010,Reiserer2015,Cirac1997,vanenk1998,Kimble2008,Wehner2018,Uphoff2016} and constitute a novel platform for many-body physics with long-range interactions~\cite{Alibart2016,Chang2018}. While most atom-photon interfaces to date operate at visible or near-infrared wavelengths ($\sim700-1000\,$nm), compatibility with telecom wavelengths ($\sim1.25-1.65$ $\mu$m) is highly desired; both for quantum communication due to low propagation loss in fiber-optic cables, and for compatibility with silicon-based photonics. Accordingly, most approaches to quantum communication require frequency conversion of single photons into the telecom window, which often results in additional noise photons and reduced efficiencies~\cite{Bock2017}. A platform combining \textit{both} long atomic coherence times and high emission bandwidth at telecom wavelengths has yet to be developed. 
 
Atom-like defects in solids~\cite{Sipahigil2016,Didos2018,Asadi2017,Kutluer2017,Zhong2017} and trapped neutral atoms~\cite{Goban2013,Thompson2013a,Tiecke2014,Chang2014,Goban2015} coupled to photonic crystal cavities hold promise to achieve such light-matter interactions. Atom-like defects in solids require no external trapping potential since they are held in the crystal field of the host solid-state environment. However, this environment has drawbacks, such as inhomogeneous broadening, phonon broadening, and spectral diffusion~\cite{Aharonovich2016,Awschalom2018}. Hence, these systems require cooling to cryogenic temperatures to reduce phonon broadening, and spectral tuning to achieve indistinguishability~\cite{Sipahigil2016}. Moreover, the atom-like defects investigated to date are either outside of the telecom window, have short coherence times, or have low emission bandwidths~\cite{Awschalom2018}.

Optically trapped atoms in free space offer the prospect of significantly improved coherence properties since inhomogeneous broadening and spectral diffusion are negligible. However, an outstanding challenge is to reliably trap an atom sufficiently close to a photonic device. Previous trapping efforts were based on evanescent fields that confine the atom near the device~\cite{Vetsch2010,Goban2012,Goban2013}, or by forming a standing wave trap via reflection from the device~\cite{Thompson2013a,Kim2018a}. Surface effects such as Van der Waals forces~\cite{McGuirk2004,Hunger2010b}, surface patch charges~\cite{Obrecht2007a}, and Casimir-Polder forces~\cite{Antezza2004,Lin2004,Obrecht2007b} complicate these approaches. Moreover, the emission wavelengths of the atomic species used to date -- rubidium and cesium -- are $\sim800$ nm, well outside the telecom window.

\begin{figure}[t]
\centering
\includegraphics[width=8.6cm]{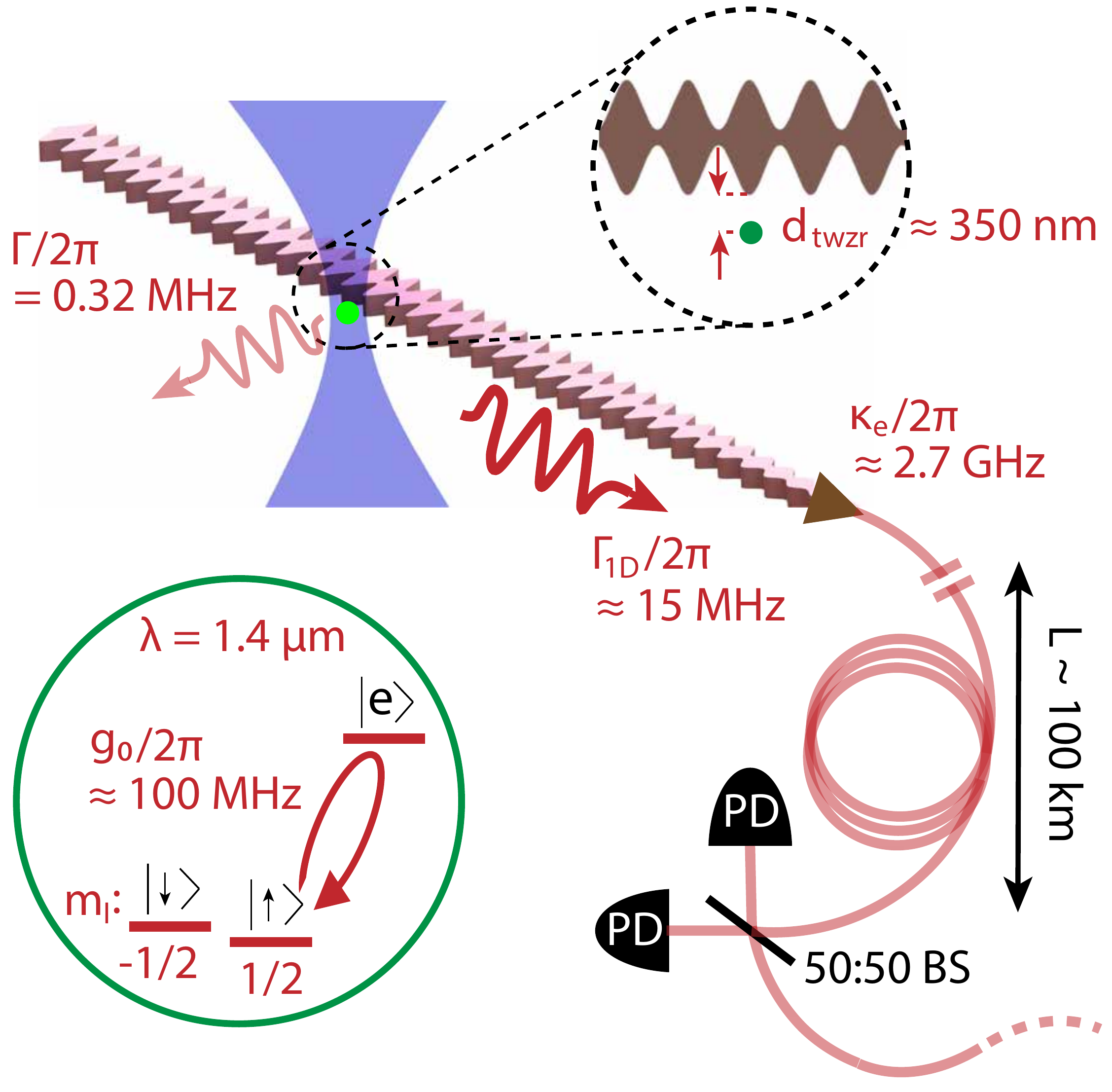}
\caption{\textbf{Schematic overview} Silicon photonic crystal cavity with an $^{171}$Yb atom trapped nearby in an optical tweezer. The minimum atom-device separation $d_\text{twzr}\approx350$ nm allowed by our approach corresponds to an atom-cavity system on a strong telecom-band transition with vacuum Rabi frequency $g_0/{2\pi}\approx100$ MHz and emission bandwidth of $\Gamma_\text{1D}/{2\pi}\approx15$ MHz, for a partially-open cavity with external coupling $\kappa_e/{2\pi}\approx2.7$ GHz and atomic free-space linewidth of $\Gamma/{2\pi}=0.32$ MHz. The nuclear spin projections $m_I$ are of the $I=1/2$ nuclear spin of $^{171}$Yb. The photon in the cavity is coupled to an optical fiber with length $L\sim100$ km. This system constitutes a node in a telecom quantum repeater in which entanglement between nodes is established by a Bell state measurement using a 50:50 beamsplitter (BS) and single-photon detectors (PD).}
\label{fig:Figure1_new}
\end{figure}

\section{Overview of the system}
We propose a platform based on short-wavelength optical tweezer trapping~\cite{Schlosser2001,Kaufman2012,Thompson2013b} to hold an Yb atom near a silicon photonic crystal in order to obtain strong atom-cavity interactions. The atomic transition is from the metastable `clock' state, and has a wavelength of $\lambda=1.4$ $\mu$m. Compared to previous work with tweezers operating at $\lambda_\text{twzr}\approx800$ nm~\cite{Tiecke2014}, we use $\lambda_\text{twzr}\approx\,$470 nm to obtain tighter focusing. Further, the use of a $\sim2\times$ longer wavelength transition results in a larger spatial extent of the evanescent cavity field. Accordingly, we propose to trap the atom without the use of reflection from the device. We use a larger distance from the device compared to previous work~\cite{Thompson2013b} (d$_\text{twzr}=350$ nm), at which surface forces are reduced by a factor of $>10$. The larger disparity between the trapping- and the telecom-transition-wavelengths in Yb enables both a five-fold increase in cooperativity and more robust atom trapping.

We focus on quantum communication as a specific application of this platform, and we envision an Yb atom coupled to a silicon nanophotonic cavity as a node in a quantum repeater network (Fig.~\ref{fig:Figure1_new}). To this end, we propose a partially-open cavity design which enables the emission of $\approx15$ MHz-bandwidth photons entangled with the nuclear spin of $^{171}$Yb that serves as a long-lived quantum memory. Further, we consider a fiber gap Fabry-P\'erot cavity in Appendix III rather than a photonic crystal, which may offer a simpler alternative, but is not compatible with on-chip silicon photonics.

We highlight the use of silicon for the photonic crystal cavity not only due to low losses but also its maturity as a fabrication technology~\cite{Almeida2004}. Robust and high-yield electronic, mechanical, and optical devices have been realized in silicon-based systems utilizing a wide array of highly developed micro- and nano-fabrication techniques. Indeed, custom silicon devices are increasingly commericially available from fabrication foundries (see e.g. Ref.~\cite{Imec}). Moreover, silicon is compatible with other photonic technologies~\cite{Miller2017} such as electro-optic~\cite{Dieterle2016,Pitanti2015} and opto-mechanical~\cite{Paraiso2015,Navarro2018} systems.

\begin{figure}[t]
\centering
\includegraphics[width=8.6cm]{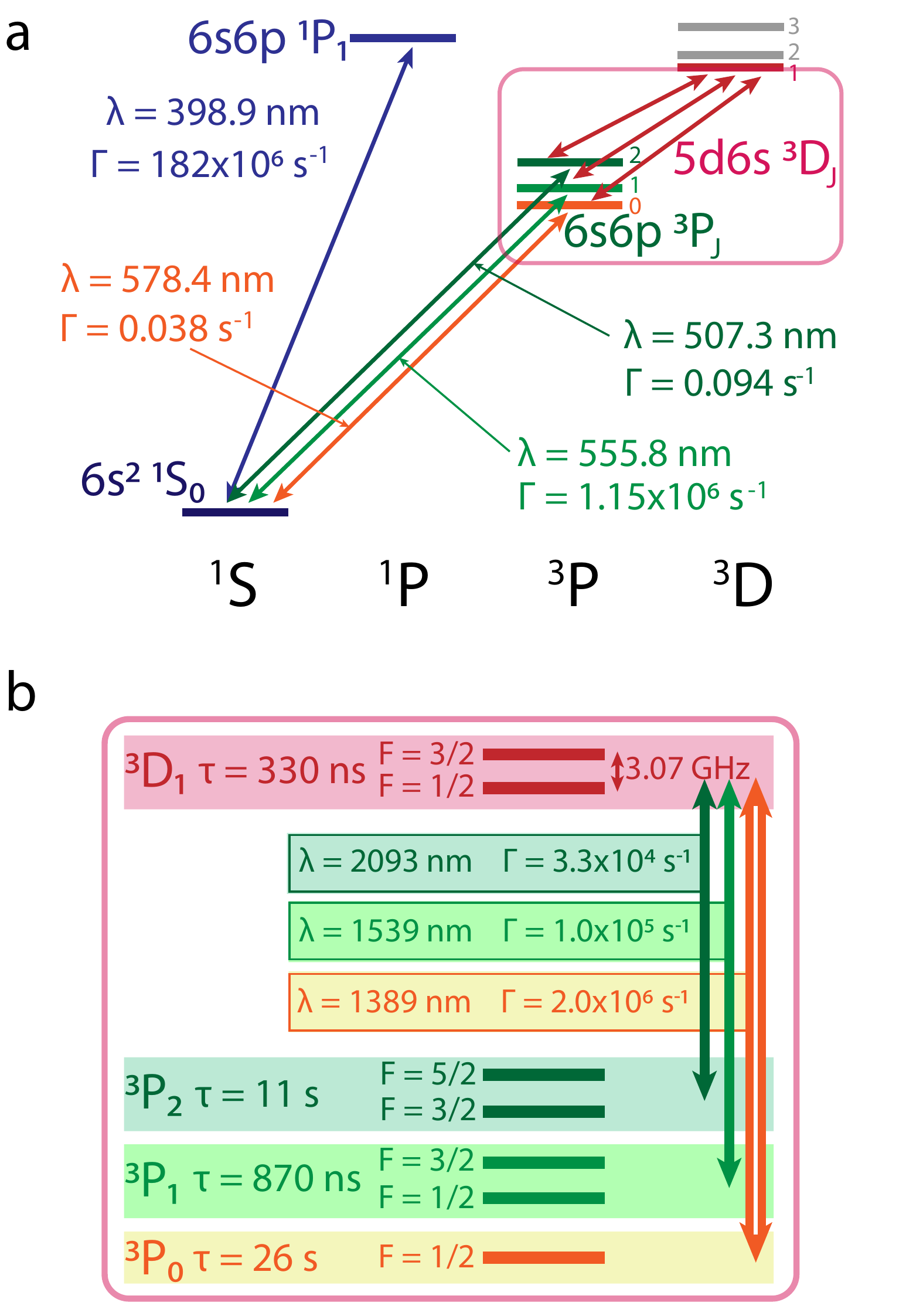}
\caption{\textbf{Level diagram of the relevant states of $^{171}$Yb} (a) Low-lying states of Yb in the singlet and triplet manifolds. The telecom transitions from the metastable 6s6p $^3$P$_\text{J}$ states to the 5d6s $^3$D$_1$ state are highlighted in the red box. (b) Zoom-in of the highlighted transitions. The nuclear spin in $^{171}$Yb is $I=1/2$, so the hyperfine states are given by $F=1/2$ when $J=0$ and $F=\{J+1/2,J-1/2\}$ when $J\geq1$. The lifetimes of the $^3$P$_\text{J}$ states, transition wavelengths, transition linewidths, as well as lifetime and hyperfine splitting of the $^3$D$_1$ state are given. We employ the transition shown with the orange double arrow.}
\label{fig:Figure1}
\end{figure}

The precise control of single atoms in this approach enables scalable extension to multiple atoms by employing recently-demonstrated techniques with tweezer arrays~\cite{Barredo2016,Endres2016,Kim2018a}. This would enable photonic coupling within an array of atoms, which could lead to a novel platform for many-body physics~\cite{Douglas2015,Hood2016}, quantum nonlinear optics~\cite{Chang2014,Goban2015}, and photon-mediated quantum gates~\cite{Welte2018}. The latter application is relevant to quantum repeaters, where deterministic two-qubit gates in each node could enhance entanglement distribution rates by realizing efficient Bell state measurements for entanglement swapping operations. Note that two-qubit gates could also be accomplished using local exchange~\cite{Daley2008,Daley2011,Kaufman2015} or Rydberg~\cite{Saffman2010} interactions.  

The strong telecom-wavelength transition of Yb is from a metastable state with lifetime $\tau\approx26$ sec (Fig.~\ref{fig:Figure1}), which is the crutial state in the optical clock transition~\cite{Ludlow2015}. We focus on the 1.4 $\mu$m ($^3$P$_0\rightarrow^3$D$_1$) transition which is shown with the orange double arrow (see Appendix I). Concerning the other transitions available, the one at 1.5 $\mu$m ($^3$P$_1\rightarrow^3$D$_1$) is hampered by the short lifetime of $^3$P$_1$, which restricts its use to more complex protocols. Finally, the 2.1 $\mu$m transition ($^3$P$_2\rightarrow^3$D$_1$) is not suitable for fiber-optic communication. However, it is an interesting candidate for free-space communication given the relatively high atmospheric transmission at this wavelength. We define the states of interest as $|\downarrow\rangle\equiv~^3\text{P}_0~|F=1/2,m_F=-1/2\rangle$, $|\uparrow\rangle\equiv~^3\text{P}_0~|F=1/2,m_F=1/2\rangle$, and $|e\rangle\equiv~^3\text{D}_1~|F=3/2,m_F=3/2\rangle$ (see Figs.~\ref{fig:Figure1_new}~\&~\ref{fig:Figure4}).

\section{Application in a quantum repeater}
Before providing a detailed description of the photonic cavity and the coupling of the atom, we briefly highlight the potential of this platform for quantum communication (see Appendix II for more details). As an example, we consider utilizing our system in a quantum repeater architecture using the Barret-Kok scheme~\cite{Barrett2005,Bernien2013}. The key parameters of our system that impact the entanglement generation rate and fidelity are summarized in Table~\ref{tab:Table1}. We describe in detail in the following sections how these values are achieved in our platform.

\begin{table}
\begin{center}
\caption{The values relevant for a quantum repeater with a photonic crystal cavity, and the sections of the text in which they are described.}
\label{tab:Table1}
\begin{tabular}{l|c|r}
\textbf{Parameter} & \textbf{Section} & \textbf{Value} \\
\hline
Bandwidth & IV & 2$\pi\times15$ MHz \\
Wavelength & II & 1.39 $\mu$m (0.35 dB/km) \\
System efficiency & IV & 0.80 \\
Memory & II, AII & $\leq26$ sec \\
Read-out fidelity & AII & $>0.99$ \\
\end{tabular}
\end{center}
\end{table} 

Our system allows a bandwidth of $\Gamma_\text{1D}=2\pi\times15$ MHz, which is sufficiently high to not limit performance of the repeater. High bandwidth emission increases the detection fidelity since the acquisition time is reduced and detection of dark counts can be mitigated. While most emitter platforms have sufficient bandwidth, many of the platforms operating in the telecom band, such as rare-earth ions in crystals, have slow emission rates. The bare linewidth of erbium (Er) ions, for instance, is $\Gamma/{2\pi}\approx14$ Hz~\cite{Bottger2006}, and thus large Purcell enhancement in high-Q nanophotonic cavities~\cite{Seidler2013,Asano2017} is required to enhance the emission rate.

The long-lived memory of the nuclear spin qubit is one of the strengths of our platform. Some atom-like defects such as nitrogen vacancy (NV) centers in diamond also have long memory~\cite{Maurer2012,Reiserer2016}, but their optical transitions are at visible wavelengths and are hampered by phonon broadening. Further, many of the solid-state systems whose optical transitions are in the telecom band have short memories~\cite{Probst2015}. In our system, the memory is assumed to be limited to the lifetime of the $^3$P$_0$ state, though care must be taken to mitigate various heating mechanisms associated with tweezer trapping. A unique feature of alkaline-earth (-like) atoms is the possibility of cooling via electronic states while preserving coherence of the nuclear spin~\cite{Daley2011}. 

\begin{figure}[t]
\centering
\includegraphics[width=8.6cm]{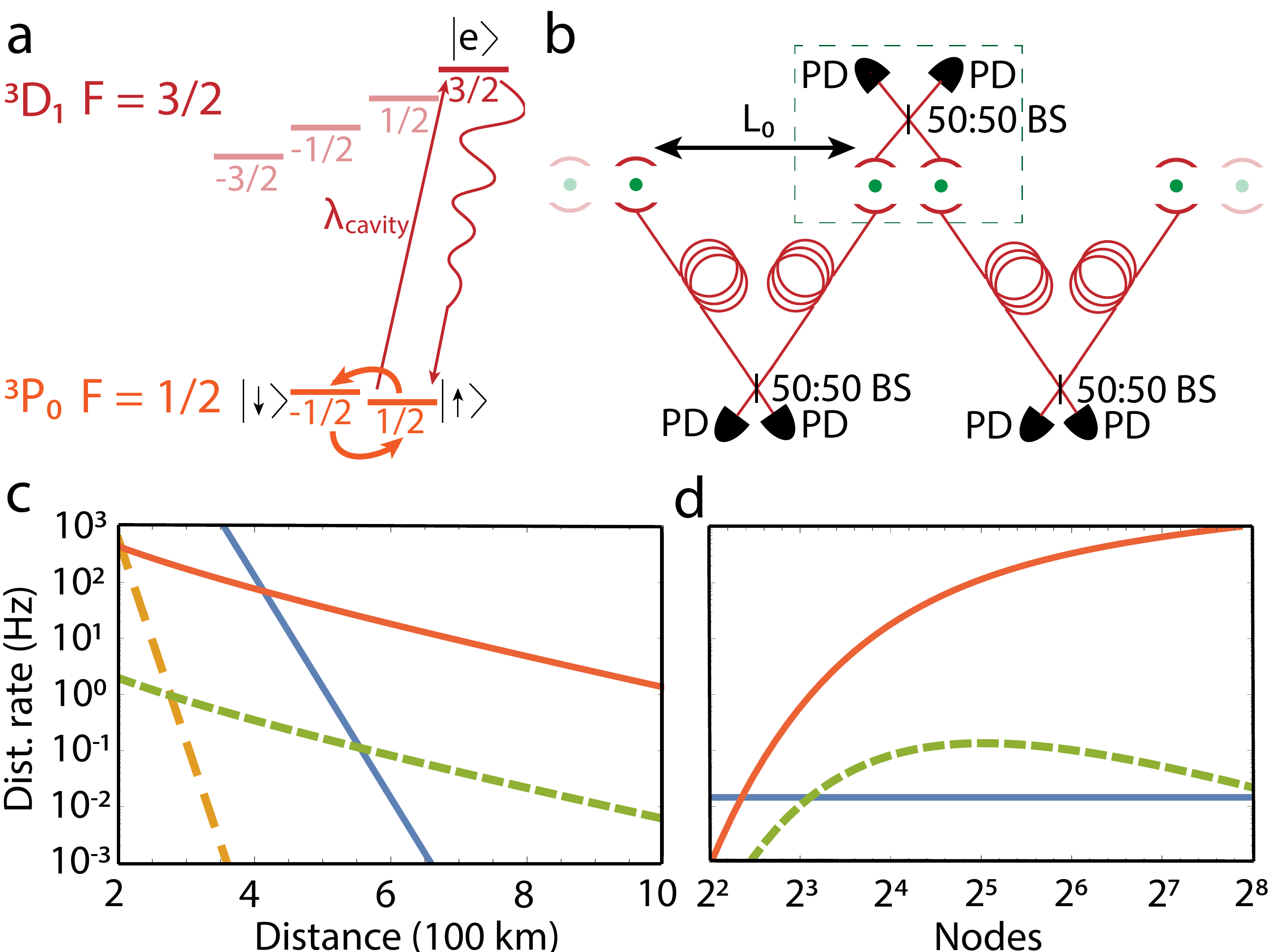}
\caption{\textbf{Spin-photon entanglement scheme and quantum repeater operation} (a) The relevant (irrelevant) hyperfine states are shown in solid (semi-transparent) colors. The cavity-enhanced transition wavelength is $\lambda_{\text{cavity}}$. (b) A single trapped $^{171}$Yb atom in a cavity is represented by a green dot inside two curved semi-circles. Local node pairs are shown in the dashed box. Node pairs are seperated by $L_0$. BS: beamsplitter, PD: single photon detector. (c) The entanglement distribution rate versus total distance via direct communication at 10 GHz for 1550 nm (solid blue) and 1390 nm (large-dashed orange) and via a quantum repeater with $2^4$ nodes with (without) local deterministic entanglement, solid red (short-dashed green). Note that the direct transmission scheme can be interpreted as the Pirandola bound for a repetition rate of 6.9 GHz~\cite{Pirandola2017}. (d) The entanglement distribution rate over 600 km versus number of nodes via direct communication at 10 GHz for 1550 nm (solid blue) and 1390 nm (large-dashed orange, not visible) and via a quantum repeater with (without) local deterministic entanglement, solid red (short-dashed green).}
\label{fig:Figure4}
\end{figure}

Another important characteristic for determining the quantum repeater performance is the system efficiency, which describes the probability that a photon emitted by the atom is acquired into the fiber network. This includes the coupling to the cavity, the extraction from the cavity into the waveguide, and the coupling to a fiber. All these values are described in Sec. IV, and the total photon system efficiency is expected to be $\eta_\text{tot}\approx0.80$.

As a concrete demonstration of the potential of this system, we calculate the entanglement distribution rate in a network as shown in Fig.~\ref{fig:Figure4}a and~\ref{fig:Figure4}b, and compare it to direct communication without repeaters (see Appendix II for analysis). For a repeater system of 16 nodes we find that the distribution rate exceeds that of direct communication with a 10 GHz single photon source for a minimum total distance of 550 km. The corresponding entanglement distribution rate is 0.1 Hz. However, when local entanglement swapping at a node can be realized using two-qubit gates rather than probabilistic photon detection-based schemes, the distribution rate could be enhanced to 25 Hz. The distribution rate versus distance for 16 nodes is shown in Fig.~\ref{fig:Figure4}c, and the rate versus number of nodes for 600 km is shown in Fig.~\ref{fig:Figure4}d. These findings indicate that this platform is a competitive quantum repeater technology in the telecom band. Note that the fiber gap Fabry-P\'erot alternative also performs well (see Appendix III).  

\section{The silicon photonic crystal cavity}
We now describe the design of the partially-open cavity, and the resulting coupling strength to an Yb atom. We consider a photonic crystal geometry based on a nanobeam with an external corrugation~\cite{Yu2014}. The sinusoidal modulation along the outer edges induces a photonic bandgap, which in turn enables the creation of a cavity via the introduction of a defect cell in the lattice to break the translational symmetry of the crystal. This enables the formation of modes localized in space around the defect region. For our chosen photonic crystal geometry, this is achieved by using a lattice constant $a_\text{mirror}=454$ nm. This is then subsequently tapered down to $a_\text{cavity}=433$ nm such that the relevant band-edge of the mirror region is tuned into the bandgap and hence establishes a cavity region. 

\begin{figure}[t]
\centering
\includegraphics[width=8.6cm]{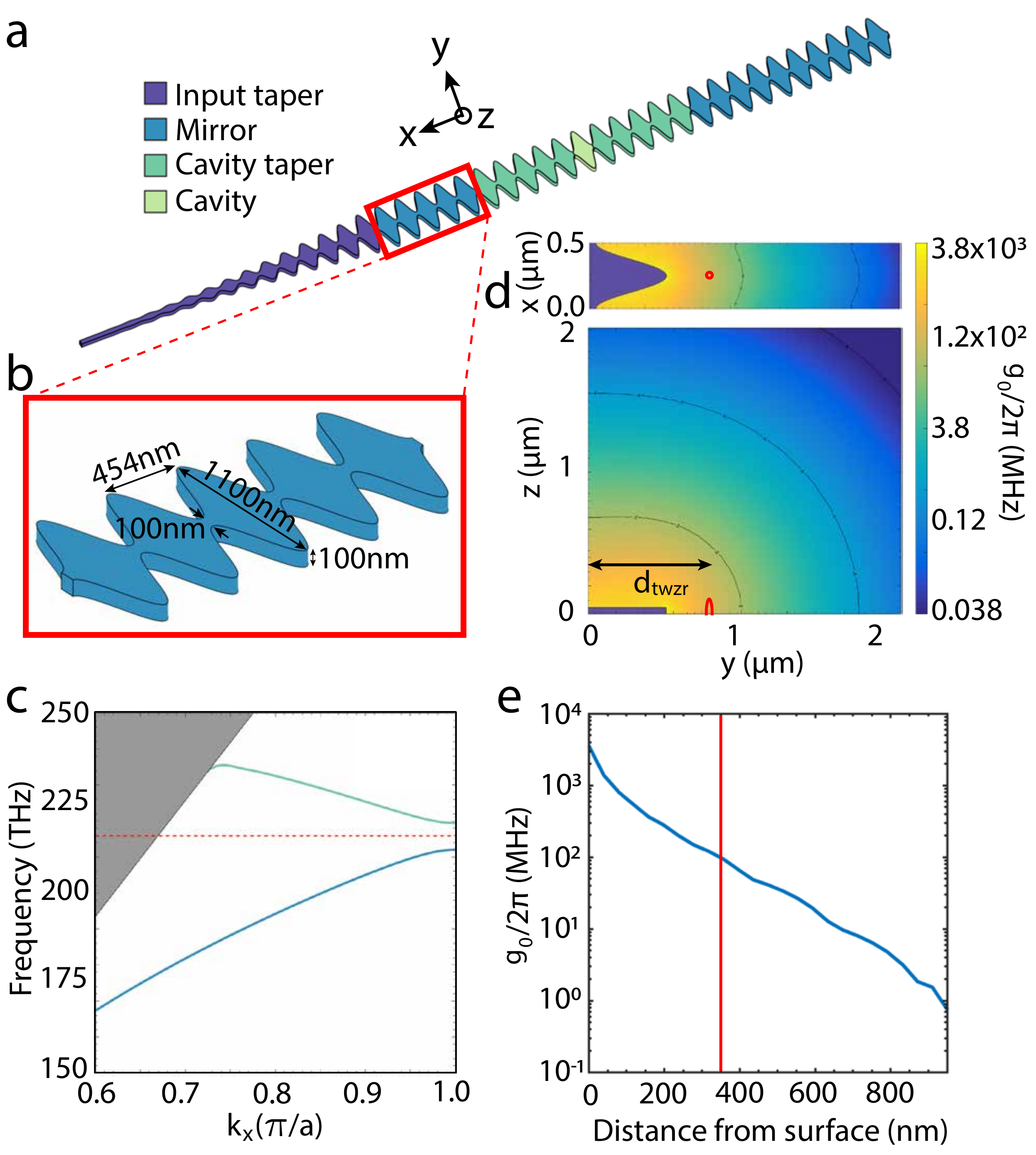}
\caption{\textbf{Design and characterization of the Si cavity} (a) Schematic of the photonic crystal cavity. The different colors show the different sections of the cavity. (b) A zoom-in of the first mirror section of the photonic cavity. (c) The TE mode band structure of the air (green) and dielectric (blue) mode. The red dashed line is the atomic resonance, and the shaded gray region is outside the light cone. (d) The coherent coupling rate $g_0$ shown on a color map as a function of distance in the $y$ and $z$ directions from the center antinode of the photonic cavity ($x=0$) in the bottom image, and the profile along the $x$ direction across the center tooth in the top image. The red ellipse and circle show the size of the atomic motional wavefunction (see Sec. V). (e) A line cut of the coherent coupling rate for $x=z=0$ in units of the distance from the surface. The position of the atom is represented by the vertical red line at $d_\text{twzr}=350$ nm.}
\label{fig:Figure2}
\end{figure}

Details of the photonic crystal geometry are shown in Fig.~\ref{fig:Figure2}a and b. The different colors show the different sections of the cavity. From left to right: purple is the input section of the cavity which transitions from a single mode waveguide to the photonic crystal geometry enabling the coupling of light in and out; medium blue is the left cavity mirror with higher transmission; dark green is the cavity taper region from the mirror cell lattice constant $a_\text{mirror}$ to the center cavity cell lattice constant $a_\text{cavity}$; light green is the central cavity unit cell; dark green is the taper to the backside mirror; and blue is the backside mirror with very high reflectivity. The tapering is done in such a way as to produce an effective quadratic potential for localized cavity photons, providing the optimal balance between localization in the plane of the device and radiation out-of-plane~\cite{Srinivasan2002,Chan2009,Chan2012}. The device thickness is chosen to be 100 nm to extend the evanescent field due to weaker confinement inside the dielectric, hence allowing for a greater distance between the photonic crystal and the atom.

This design has a radiation limited quality factor of $3.5\times 10^6$ in simulations. However, we anticipate that the quality factor will be limited by intrinsic fabrication imperfections to $Q_i\leq7\times10^5$, whose corresponding intrinsic cavity linewidth is $\kappa_i=2\pi\times300$ MHz~\cite{Fang2017}. We design the cavity to be partially open on one side (as in Fig.~\ref{fig:Figure2}a) to efficiently extract the cavity photons~\cite{Gallego2016,Gallego2018,Welte2018}. Specifically, we consider $5$ mirror cells on the front mirror and $10$ on the back mirror. The collection efficiency $\eta_\text{coll}$ of extracting the photon into the waveguide mode is given by $1-Q_e/Q_i$, where the subscript `$e$' denotes external coupling. In our design we have chosen a modest $Q_e=8\times10^4$, for which $\eta_\text{coll}=0.89$ and $\kappa_e=2\pi\times2.7$ GHz. 

We now consider the photonic mode profile. The transverse electric (TE) photonic band structure containing a bandgap centered on the atomic transition at $\lambda=1388.8$ nm (215.9 THz) is shown in Fig.~\ref{fig:Figure2}c, and the evanescent field profile of the dielectric mode is shown in Fig.~\ref{fig:Figure2}d. The blue ellipse and circle show the $1/e^2$ size of the atomic motional wavefunction, to be discussed in the next section. A line cut of the coherent coupling rate is shown in Fig.~\ref{fig:Figure2}e for $x=z=0$ versus the distance from the surface, and we find an exponential length scale of $\lambda_\text{Si}\approx170$ nm. This is consistent with estimates for the mode dispersion based on the band structure (Fig.~\ref{fig:Figure2}c). The position of the atom is represented by the vertical red line. 

At the location of the maximal evanescent field, the electric field per photon is $\text{E}_\text{cavity}=2.6\times10^5$ V/m.  The vacuum Rabi frequency (i.e. coherent coupling rate) can then be calculated to be $g_0=2\pi\times3.8$ GHz at this location, and $g_0=2\pi\times100$ MHz at the chosen location of the atom $d_\text{twzr}=350$ nm, as explained in Sec. V. The single-atom cooperativity defined here as $C_0=4g_0^2/\kappa\Gamma$ between a Yb atom and the silicon photonic crystal cavity can be estimated using the cavity linewidth $\kappa=\kappa_i+\kappa_e$ and the atomic linewidth $\Gamma=2\pi\times0.32$ MHz. This corresponds to a cooperativity of $C_0=47$. The Purcell-enhanced emission rate $\Gamma_\text{1D}$ is given by the Purcell factor $P=C_0$ and the atomic decay rate $\Gamma$ as $\Gamma_\text{1D}=P\Gamma$, which for this system gives $\Gamma_\text{1D}/2\pi=15$ MHz. 

The probability of spontaneously emitting a photon into the cavity mode $P_\text{cavity}$ is given by $C_0/(C_0+1)$, which is 0.98. Thus, the total efficiency of extracting the photon from the atom into the waveguide mode is given by $\eta_\text{ext}=P_\text{cavity}\cdot\eta_\text{coll}$, which is $0.87$. We design the photonic cavity to taper to a nanobeam waveguide which can then be coupled to an optical fiber using a microlens or adiabatic coupler. Efficiencies for the latter are $\eta_\text{AC}\approx0.95$~\cite{Tiecke2015}. These parameters result in a total system efficiency of $\approx0.80$.

\section{Trapping a single Yb atom near a photonic crystal}
We now show that an Yb atom can be trapped close to the photonic crystal using only a tightly-focused optical tweezer.

\subsection{Analysis of the optical tweezer trap}
Single-atom detection and addressing of alkaline earth (-like) atoms is a growing area of research interest. Quantum gas microscopy of Yb has been demonstrated~\cite{Yamamoto2016}, and large two-dimensional tweezer arrays of Yb~\cite{Saskin2018} and strontium (Sr)~\cite{Cooper2018,Norcia2018} have recently been reported. Further, cooling of single alkaline-earth (-like) atoms close to the motional ground state of an optical tweezer has recently been demonstrated for Sr~\cite{Cooper2018,Norcia2018}, and cooling of alkali atoms optically trapped $\approx300$ nm from a room-temperature surface has recently been observed~\cite{Meng2018}.

For a tweezer wavelength of $\lambda_\text{twzr}\approx$473 nm (depending on the tweezer polarization~\cite{Cooper2018}), there is a `magic' wavelength for which the polarizability of $^1$S$_0$ and $^3$P$_1$ are identical~\cite{Beloy2012,Tang2017}. This is particularly useful for cooling the atom in the tweezer~\cite{Cooper2018,Norcia2018}. Coincidentally, the polarizability of $^3$P$_0$ is also similar~\cite{Scazza2015t}. Moreover, the polarizability at this wavelength is large, which allows deep traps to further mitigate surface forces. As such, we propose to use the $\approx473$ nm wavelength for generating tightly-focused optical tweezers, although easily accessable wavelengths such as 532 nm are an alternate as they have been used in a similar magic configuration~\cite{Yamamoto2016,Saskin2018}. 

In order to further understand the design constraints and tweezer trap properties, we describe here the polarizability at the trapping wavelength $\lambda_\text{twzr}$, and the tweezer waist that can be generated with numerical aperture NA$\approx0.7$ objective~\cite{Yamamoto2016}. The polarizability of $^1$S$_0$ and $^3$P$_1$ at $\lambda_\text{twzr}$ is $\alpha=-18$ Hz/(W/cm$^2$)~\cite{Scazza2015t}. For an objective of NA$\approx0.70$, the $1/e^2$ waist radius of a tweezer that can be generated with this wavelength is $w_\text{twzr}\approx330$ nm, and the corresponding Rayleigh range is $R_\text{twzr}\approx730$ nm. An optical power of 1.0 mW is required for a trap depth of $U_\text{twzr}\approx0.5$ mK, and the trapping frequencies are $\omega_R\approx2\pi\times150$ kHz and $\omega_z\approx2\pi\times48$ kHz. The polarizability of the $^3$D$_1$ state is not well known, but it is only populated during a $\pi$-pulse for photon-spin entanglement and during read-out (see Appendix II). To mitigate deleterious effects from a polarizability mismatch on the $^3$P$_0\leftrightarrow^3$D$_1$ transition, we propose to switch the trap off during excitation~\cite{Tiecke2014}. Atom survival probability in the absence of a trap is known to be high for times of several $\mu$s~\cite{Bernien2017}, which is much longer than the Purcell enhanced emission timescale $\Gamma_{1D}^{-1}=11$ ns.

The temperature of a typical Yb magneto-optical trap (MOT) operating on the $^1$S$_0$ - $^3$P$_1$ transition is $<10$ $\mu$K~\cite{Scazza2015t,Yamamoto2016}. This temperature would correspond to $n_R=0-1$ motional quanta in the radial direction and $n_z=2-3$ motional quanta in the axial direction, and further cooling in a tweezer has been demonstrated with Sr~\cite{Cooper2018,Norcia2018}. Such conditions lead to thermal $1/e^2$ atomic wavefunction radii of $\sigma_R\approx30$ nm in the radial direction and $\sigma_z\approx80$ nm in the axial direction. The circles in Figs.~\ref{fig:Figure2}d are meant to roughly represent the size of the atomic wavefunction and illustrate how small it is compared to the mode profile of the cavity field. Note that the nodal spacing of the cavity is larger here compared to previous work~\cite{Thompson2013a} because of the longer wavelength, and thus the evanescent field coupling is expected to be more homogeneous over the tweezer trap volume.

\begin{figure}[t]
\centering
\includegraphics[width=8.6cm]{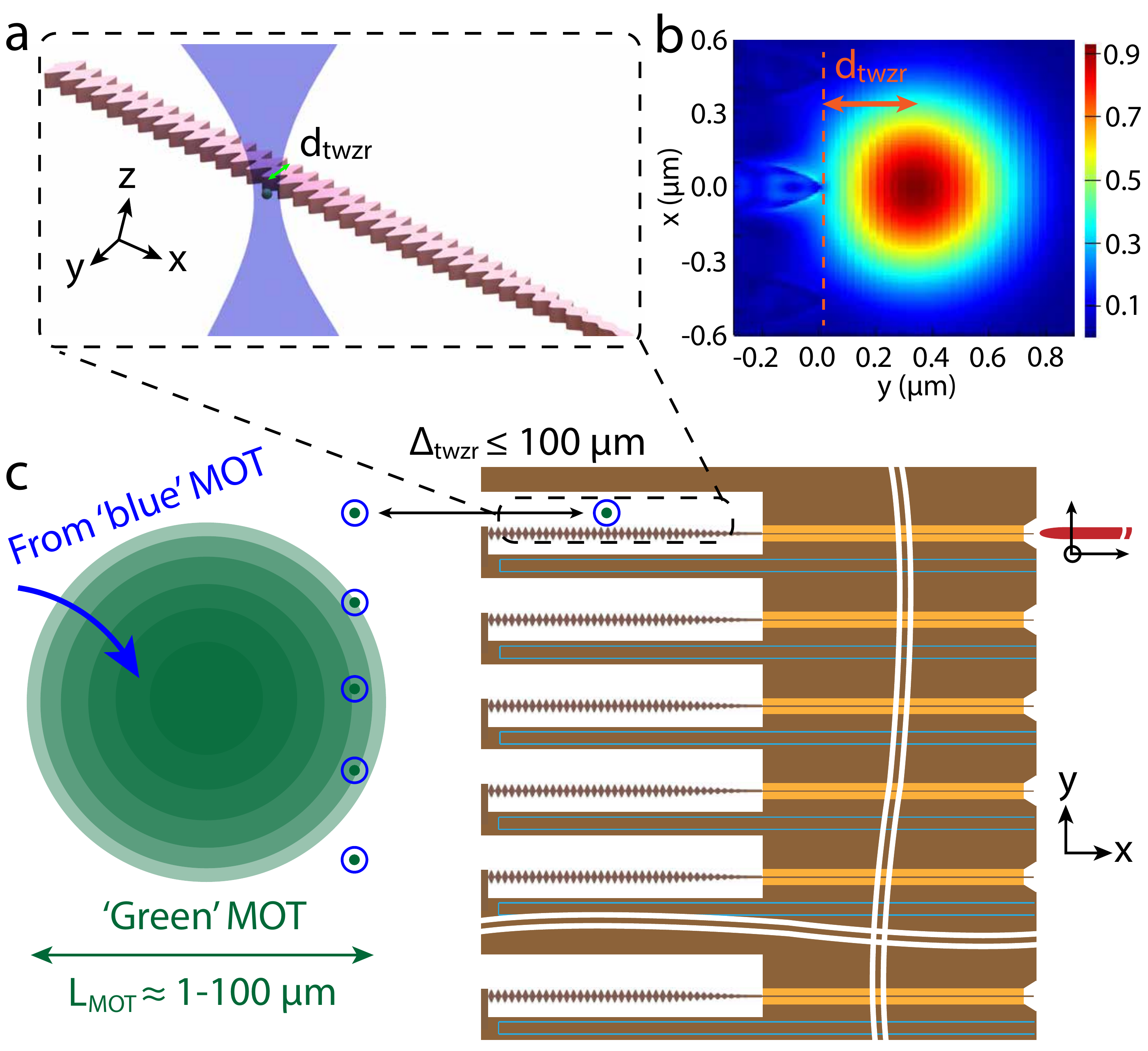}
\caption{\textbf{Coupling a Yb atom to a silicon photonic crystal cavity via an optical tweezer trap} (a) A zoom-in illustrating the atom in a tweezer near the device at a distance $d_\text{twzr}$. (b) The electic field magnitude of the tweezer trap in arbitrary units at a distance of $d_\text{twzr}=350$ nm from the edge of the device. (c) Schematic of the silicon chip. The silicon top layer is brown, and the insulator layer below is orange. The blue microstrips on the chip are used for Ohmic temperature control and applying RF magnetic fields. The lensed fiber on the right can be coupled to any cavity using a 3D translation stage. Wavy white lines indicate cuts to show the entire chip. The green circle on the left is the Yb MOT, and atoms can be shuffled between it and the device using optical tweezers. This architecture allows simultaneous operation of multiple Yb-cavity nodes on a single device.}
\label{fig:Figure3T}
\end{figure}

\subsection{Atom trapping and imaging near the photonic crystal}
We consider a tweezer focused at a distance d$_\text{twzr}=350$ nm from the photonic crystal, as shown in Fig.~\ref{fig:Figure3T}a and \ref{fig:Figure3T}b. This was chosen to be slightly larger than the waist of the tweezer $w_\text{twzr}\approx330$ nm to minimize the impact of scattered fields from the photonic crystal. Further, we propose to use the chip geometry shown in Fig.~\ref{fig:Figure3T}c and discussed in the next subsection, which allows the tweezer to be translated with respect to the cavities on the chip and the MOT~\cite{Thompson2013a} using an acousto-optic deflector or spatial light modulator. This is advantageous since atomic flux on the device is known to have detrimental effects on photonic structures~\cite{Ritter2015,Ritter2016}.

A significant difference between alkali- and alkaline-earth(-like) atoms is the size and temperature of their MOTs. For alkalis, typically only one atomic transition is used for laser cooling and optical molasses. Conversely, alkaline-earth(-like) atoms are typically laser cooled in two stages: the broad $^1$S$_0\leftrightarrow^1$P$_1$ transition for initial loading and the narrow $^1$S$_0\leftrightarrow^3$P$_1$ transition for cooling to $<10$ $\mu$K~\cite{Scazza2015t,Yamamoto2016} as mentioned above. Further, the narrow linewidth of the cooling transition in the second stage allows for flexibility in the MOT size and position by adjusting the magnetic field gradient and offset. Hence, a small and cold second-stage `green' MOT could be moved to $\approx100$ $\mu$m off the silicon chip using magnetic fields, which is within the field of view of typical microscope objectives. This enables easy tweezer transport with acousto-optic deflectors~\cite{Endres2016} (see Fig.~\ref{fig:Figure3T}), and does not require translation of the objective. The large first-stage `blue' MOT could be millimeters from the chip at all times. This approach will reduce atomic flux onto the device compared to previous work~\cite{Thompson2013a,Goban2013}, and also mitigates the detrimental effect of the photonic structure on the MOT.

Surface effects on the atom are substantially mitigated in our approach compared to previous work~\cite{Thompson2013a}. Operating at 350 nm from the surface instead of 200 nm decreases the surface potential to $\approx13$\%, and surface force to $\approx7$\% to that of previous work~\cite{Thompson2013a} (see Supplementary Material in Ref.~\cite{Thompson2013a}, and Ref.~\cite{Derevianko2010} for electric dipole polarizabilities at imaginary frequencies used in this calculation). In addition, the external corregation used in our photonic crystal design further reduces the effective surface area interacting with the atom where only the surface areas at the antinodes (see Fig.~\ref{fig:Figure1_new} and~\ref{fig:Figure3T}b) contribute significantly to the surface force. We can compare this surface force with the dipole force in the tweezer potential. We expect the surface force to be 100's of kHz/$\mu$m or less, while the maximum dipole force for a tweezer as described above with 1 mK depth is $\approx50$ MHz/$\mu$m, and the use of deeper traps is possible.

Now we consider absorption of the tweezer light by the silicon nanocavity. Absorption can have two deleterious effects: it can cause heating of the structure that will alter the cavity properties, and it can generate free-carriers which increase the optical absorption in the telecom band~\cite{Barclay2005,Johnson2006}. The absorption coefficient of silicon at a wavelength of $\approx470$ nm is $\alpha_\text{Si}\approx2\times10^{4}$ cm$^{-1}$~\cite{Bucher1994}, which means that $\approx18\%$ of incident tweezer light will be absorbed by our cavity of 100 nm thickness. For a 1 mW tweezer of waist $w_\text{twzr}=330$ nm at a distance from the surface of $d_\text{twzr}=350$ nm, we estimate that $<3$ $\mu$W is incident on the device, and thus $<530$ nW is absorbed. Given the thermal conductivity of silicon and the dimensions of the proposed nanocavity we expect a $\approx0.3$ K temperature difference in the vicinity of the tweezer relative to the chip which serves as a thermal reservoir.  The temperature dependence of the index of refraction causes a shift in the cavity resonance frequency at the $10^{-5}$ /K level~\cite{Barclay2005}. Hence, we expect $\approx3\times10^{-6}$ fractional shift in the cavity resonance, which is much less than $1/Q_e>10^{-5}$ for the cavity.

To mitigate the effect of free-carrier generation in the silicon nanocavity, we propose to switch the tweezer off during the telecom-photon emission phases, such as spin-photon entanglement and read-out. The weak tweezer illumination onto the device will cause free electron-hole pairs which will decay within 10's of ns~\cite{Barclay2005,Johnson2006}. This is fast compared to the allowed free expansion time of the atom with high probability re-trapping (several $\mu$s), so waiting for 100's of ns to ensure electron-hole pair decay is feasible. Since the recoil energy of a telecom photon is low, we expect heating of the atom during these pulses to be sufficiently small.

Coherent scattering from the nanocavity must also be considered because it could alter the trapping potential. We calculate the trapping potential in the presence of scattering from the silicon device using a finite-difference time-domain simulation. We simulate a gaussian beam with a waist of 330 nm focused at a distance of $d_\text{twzr}=350$ nm from the edge of the photonic crystal cavity. The results are shown in Fig.~\ref{fig:Figure3T}b, where the magnitude of the electric field is shown on a relative scale. The trap perturbations are below $|\text{E}_\text{twzr}|\approx0.2$ ($I_\text{twzr}\approx0.1$), and they occur only at distances of $\geq d_\text{twzr}$ from the center of the tweezer. This effect is negligible, particularly for a cold atom.

The ability to have many photonic cavities per chip facilitates fabrication error and device degradation to be overcome, but also opens the possibility to couple an array of atoms in optical tweezers~\cite{Endres2016} to an array of cavities. This would enable multi-qubit repeater nodes as discussed in Sec. II, III, and Appendix II. An array of cavities can be fabricated such that their separation is as small as $\approx10$ $\mu$m  (as in Fig.~\ref{fig:Figure3T}c). The cavities are coupled to a microlens coupler or adiabatic coupler, and the fiber could be aligned and coupled to any individual cavity on the chip using a three-axis stage as shown in Fig.~\ref{fig:Figure3T}c.  Each cavity is independently temperature controlled with its own tungsten heating strip (light blue lines in Fig.~\ref{fig:Figure3T}c), and thus each cavity in the array could be individually tuned into resonance with the atomic transition~\cite{Kim2016}. This strip is also used for applying radio-frequency (RF) pulses to control the nuclear spin (Appendix II).

\section{Conclusion and outlook}
We have illustrated that silicon-based photonics combined with tweezer-trapped $^{171}$Yb atoms are a candidate system for quantum optics in the telecom-band. The strong $1.4$ $\mu$m transition from the $^3$P$_0$ `clock' state of Yb enables 15 MHz-bandwidth emission when coupled to a silicon photonic cavity, and the nuclear spin allows for a coherent quantum memory. Further, we propose a simple and robust trapping protocol which enables atoms to be coupled to silicon nanophotonics, and even facilitates an array of atoms coupled to an array of cavities. Furthermore, we illustrate the potential of our platform for quantum repeaters.

Moreover, there are several other applications of this system which include quantum photonic circuits, novel platforms for long-range interactions, and many-body physics~\cite{Douglas2015}. Our system may enable the direct integration of neutral-atom quantum computers into a quantum network, in which quantum gates can be performed using local exchange~\cite{Daley2011,Kaufman2015} or Rydberg~\cite{Saffman2010} interactions. Further, this system opens the possibility to implement an optical clock network~\cite{Komar2014} by using the $^1$S$_0\leftrightarrow ^3$P$_0$ optical clock qubit~\cite{Ludlow2015}.

Alternative cavity designs based on free-space optics could offer different possibilities, and we consider a fiber Fabry-P\'erot cavity in Appendix III. Finally we note that a similar telecom-wavelength atomic system could be created with Yb$^+$ ions, where a group of transitions from a metastable state has convenient wavelengths of 1450 and 1650 nm. We analyze a Yb$^+$ ion coupled with a fiber Fabry-P\'erot cavity in Appendix IV, but we note that ion trapping near dielectric materials poses other technical challenges.

\section*{Acknowledgements}
We acknowledge Mohammad Mirhosseini, Alexandre Cooper-Roy, and Matthew D. Shaw for useful discussions, and Hengjiang (Jared) Ren with help with COMSOL simulations. We also ackowledge Hannes Bernien and Jeff Thompson for critical reading of the manuscript. JPC acknowledges support from the Caltech PMA Division for post-doctoral fellowship funding, AS acknowledges support from the Caltech IQIM for post-doctoral fellowship funding, and NS acknowledges funding by the Alliance for Quantum Technologies' Intelligent Quantum Networks and Technologies (INQNET) research program. We acknowledge funding provided by the Institute for Quantum Information and Matter, an NSF Physics Frontiers Center (NSF Grant PHY-1733907). This work was also supported by the NSF CAREER award, the Sloan Foundation, and by the NASA/JPL President's and Director's Fund. 

\setcounter{section}{0}  
\renewcommand{\thesection}{APPENDIX~\Roman{section}}

\section{Dipole matrix elements and polarization considerations}~\label{sec:Dipole}
In order to quantify the coupling to a cavity we must calculate the dipole matrix element of the desired atomic transition: $^3\text{P}_0~|F=1/2,m_F=1/2\rangle \leftrightarrow ~^3\text{D}_1~|F=3/2,m_F=3/2\rangle$ (see Appendix II for detailed quantum repeater scheme). The $^3$P$_0\leftrightarrow^3$D$_1$ transition has been carefully measured because of its relevance to Yb clock precision~\cite{Beloy2012}. Given that $\Gamma_{^3\text{D}_1\rightarrow^3\text{P}_0}=2\times10^6$ s$^{-1}$, we arrive at $\langle~F=1/2~m_F=1/2||e\hat{\textbf{r}}_q||F^{\prime}=3/2~m_{F^{\prime}}=3/2\rangle=1.38\times10^{-29}$ C-m, or 1.63 a$_0$-e, in which the polarization is taken to be $q=+1$ ($\sigma^+$). However, purely circular polarization cannot be supported by the modes of the photonic structure (see Sec. IV), and so the effective dipole matrix element is reduced by $\sqrt{2}$ upon decomposing $\sigma^+$ into a combination of linear polarizations. The `quantization axis' is assumed to be determined by the electric field of the optical tweezer $\text{E}_\text{twzr}$ (see Sec. IV), and the external magnetic field $\text{B}_\text{ext}$ (see Appendix II) is assumed to be parallel. To drive the $\sigma^+$ transition, we require the electric field of the mode in the cavity $\text{E}_\text{cavity}$ to be perpendicular to the quantization field axis $\text{E}_\text{cavity}\perp\text{E}_\text{twzr},~\text{B}_\text{ext}$.

\section{Quantum repeater implementation}
In this section we describe a quantum repeater based on a network of $^{171}$Yb atoms coupled to photonic crystal cavities. The repeater involves dividing a long channel of length $L$ into $2^n$ elementary links of length $L_0$ that are connected by nodes which feature a pair of trapped atoms. The integer $n$ is often referred to as the number of nesting levels \cite{Sangouard2011}. Atom-atom entanglement that spans $L$ is achieved by entangling atoms that are separated by $L_0$ and by performing entanglement swapping between atoms that are located at each node. Fig. \ref{fig:Figure4}b depicts two elementary links and entanglement swapping at one node. We consider the scheme of Barrett and Kok ~\cite{Barrett2005,Bernien2013} for generation of remote spin-spin entanglement and then calculate the rate of distribution of a Bell state.

\subsection{Level scheme and protocol}
We consider the three-levels of $^{171}$Yb shown in Fig.~\ref{fig:Figure4}a using solid colors. These consist of an excited $m_F=3/2$ Zeeman level of the $^3$D$_1$ manifold and a pair of $m_F = \pm 1/2$ nuclear spin levels of the $^3$P$_0$ manifold that form the ground level. The $^3$P$_0$ state can be populated from the $^1$S$_0$ ground state by using the `clock' transition~\cite{Ludlow2015}, or by multi-photon processes ~\cite{Barker2016}. The $m_F=3/2$ and $m_F=1/2$ Zeeman levels of the $^3$D$_1$ level are separated by 0.47 MHz/G, which is much larger than the 752 Hz/G splitting of the $m_F=\pm1/2$ levels of $^3$P$_0$.

The first step of the repeater is to generate spin-photon entanglement and then, using two-photon detection, spin-spin entanglement between atoms that are separated by one elementary link. We begin by preparing each atom in an equal superposition of the $m_F =\pm 1/2$ nuclear states of the $^3$P$_0$ ground level: $1/\sqrt{2}( \ket{\uparrow}+\ket{\downarrow})$. This is accomplished by a RF field despite the relatively small gyromagnetic ratio of the spin $\gamma_N/{2\pi}=752$ Hz/G. Nonetheless, this favorably results in a weak coupling of the spin to the environment~\cite{Boyd2007,Gorshkov2009}. We propose to split these states with a magnetic field of $\text{B}_\text{ext}=200$ G. With the fabricated micro-stripline resonator~\cite{Robledo2011a} described in Sec. V, we expect Rabi frequencies of tens of kilohertz using tens of Watts of RF power. This allows $\pi$-pulses to be performed on timescales that are much less than the time to establish entanglement over an elementary link (see Subsec. B).

Next, each atom is excited by a short laser pulse that is resonant with the $\ket{\uparrow} \rightarrow \ket{e}$ transition, as indicated by $\lambda_{\textrm{cavity}}$ in Fig. \ref{fig:Figure4}a. As described in Sec. V, this transition is strongly-coupled to the cavity, and thus the resultant spontaneous emission locally entangles the spin and photon number in the Bell state $1/\sqrt{2}( \ket{\uparrow,1}+\ket{\downarrow,0})$, in which 1(0) represents the presence (absence) of an emitted photon. 

The photons that are emitted by each atom are directed to a beam splitter which is located half-way between nodes, see Fig.~\ref{fig:Figure4}b. If the photons that are emitted by each atom are indistinguishable, detection of one photon after the beam splitter heralds spin-spin entanglement or, due to potential loss and imperfections, a spin-spin product state between each atom ~\cite{Barrett2005,Bernien2013}. To avoid the latter, a $\pi$-pulse inverts the $m_F = \pm 1/2$ spins and the transition is optically excited a second time. The detection of a photon in both rounds heralds the creation of a spin-spin Bell state between each atom $1/\sqrt{2}( \ket{\uparrow,\downarrow}_{AB} \pm \ket{\downarrow,\uparrow}_{AB})$, in which the relative phase is defined according to whether both photons were detected on the same or different output ports of the beam splitter.

Entanglement swapping is accomplished by performing a similar procedure as to generate heralded entanglement-- optical excitation, single photon detection, spin flip, and detection of a second photon. This procedure limits the swapping efficiency to at most 50\%~\cite{Calsamiglia2001}. A deterministic swapping process, which allows improved scaling, could be achieved by trapping two atoms using two tweezers within a single cavity and exploiting photon-mediated deterministic intracavity gates ~\cite{Douglas2015}.

A limitation of this process will be the isolation of the $\ket{\uparrow} \rightarrow \ket{e}$ transition relative to the other transitions in the $^3$D$_1$ state (see Fig.~\ref{fig:Figure4}a). The splitting between the $m_F=3/2$ and $m_F=1/2$ states is 470 kHz/G, and this must be compared to the Purcell-enhanced linewidth $\Gamma_{1D}$. We choose a field of $\text{B}_\text{ext}=200$ G for which $\Delta=2\pi\times93$ MHz. For $\Gamma_{1D}=2\pi\times15$ MHz and Rabi frequency $\Omega=\Gamma_{1D}$, the off-resonant scattering rate is $\Gamma_\text{SC}=2\pi\times600$ kHz. The read-out fidelity is assumed to be $\mathcal{F}_\text{RO}=1-\Gamma_\text{SC}/\Gamma_{1D}$, which is $>0.99$. Note that this value is even slightly improved when considering the Clebsch-Gordan coefficients for the different pathways. 

\subsection{Entanglement distribution rate}
We quantify the distribution rate of a Bell state using our repeater scheme, showing that it outperforms an approach based on the direct transmission of photons. We denote the success probability for an atom to emit a photon into a single mode fiber (e.g. system efficiency) to be $p$, which includes the probability to prepare the initial state, the spontaneous emission of a photon into the cavity mode, and the coupling into a fiber. The probability of the two-photon measurement at the center of the elementary link is given by $P_0= \frac{1}{2} p^2 {\eta_t}^2 {\eta_d}^2$ in which ${\eta_t}=e^{-L_0/(2 L_\text{att})}$ is the fiber transmission with attenuation length $L_\text{att}=12$ km. This corresponds to losses of 0.35 dB/km at 1.4 $\mu$m using hydrogen-aged single mode fiber, which has been recently deployed for modern infrastructure (see, e.g., Corning SMF-28e for ITU-T~G.562D standards~\cite{Napoli2017}). 

The spin-spin entanglement creation step is repeated at time intervals of the communication time $L_0/c$, in which $c=2 \times 10^8$ m/s is the speed of light in fiber. Thus, the average time to produce entanglement that spans an elementary link is $T_{L_0}= \frac{L_0}{c} \frac{1}{P_0}$. Using the beam splitter approach depicted in Fig.~\ref{fig:Figure4}b, the efficiency of the entanglement swapping operation is $P_s= \frac{1}{2} p^2 {\eta_d}^2$, while a deterministic gate allows $P_s=1$, assuming the gate fidelity is unity. Therefore, the total time for the distribution of an entangled pair over distance 2$L_0$ is given by $T_{2L_0}= \frac{3}{2} \frac{L_0}{c} \frac{1}{P_0 P_s}$, and the average time to distribute an entangled pair over distance $L$ is
$T_{L}\approx \Big(\frac{3}{2} \Big)^{n-1} \frac{L_0}{c} \frac{1}{P_0 (P_s)^n}$. The factor of 3/2 arises because entanglement has to be created over two links before the swapping is performed~\cite{Sangouard2009,Sangouard2011}. 

For the discussion in Section III we assume $p=0.8$, which is given by the $\eta_{\textrm{ext}}\cdot\eta_{\textrm{taper}}\cdot\eta_\text{AC}$ estimated in Sec. IV, a detection efficiency of 0.9, and that the lifetime of the $^3$P$_0$ level is much longer than the distribution time. This detection efficiency is straightforwardly achieved using superconducting nanowires, which have been demonstrated at 1.5 $\mu$m~\cite{Marsili2013}. Note that the spontaneous Purcell-enhanced emission time from $^3$D$_1$ is $1/\Gamma_\text{1D}\approx11$ ns, which is negligible compared to the time to distribute entanglement over an elementary link. Figs.~\ref{fig:Figure4}c and~\ref{fig:Figure4}d show the repeater performance based on this analysis, plotted versus total distance for $2^4$ nodes (\ref{fig:Figure4}c) and versus number of nodes for a total distance of 600 km (\ref{fig:Figure4}d).

\section{Coupling Yb to a fiber gap Fabry-P\'erot cavity}
In this section, we discuss an alternative approach to efficiently interface a single atom with a single telecom photon in a fiber using fiber based Fabry-P\'erot (FP) resonator~\cite{Steinmetz2006,Colombe2007,Hunger2010}. While fiber based Fabry-P\'erot resonators have significantly larger mode volumes ($\sim1000 \lambda^3$) that result in reduced vacuum Rabi frequencies, we show that they can achieve high enough cooperativities necessary to extract single photons directly into a telecom fiber with sufficiently high efficiency. 

\begin{figure}[t]
\centering
\includegraphics[width=8.6cm]{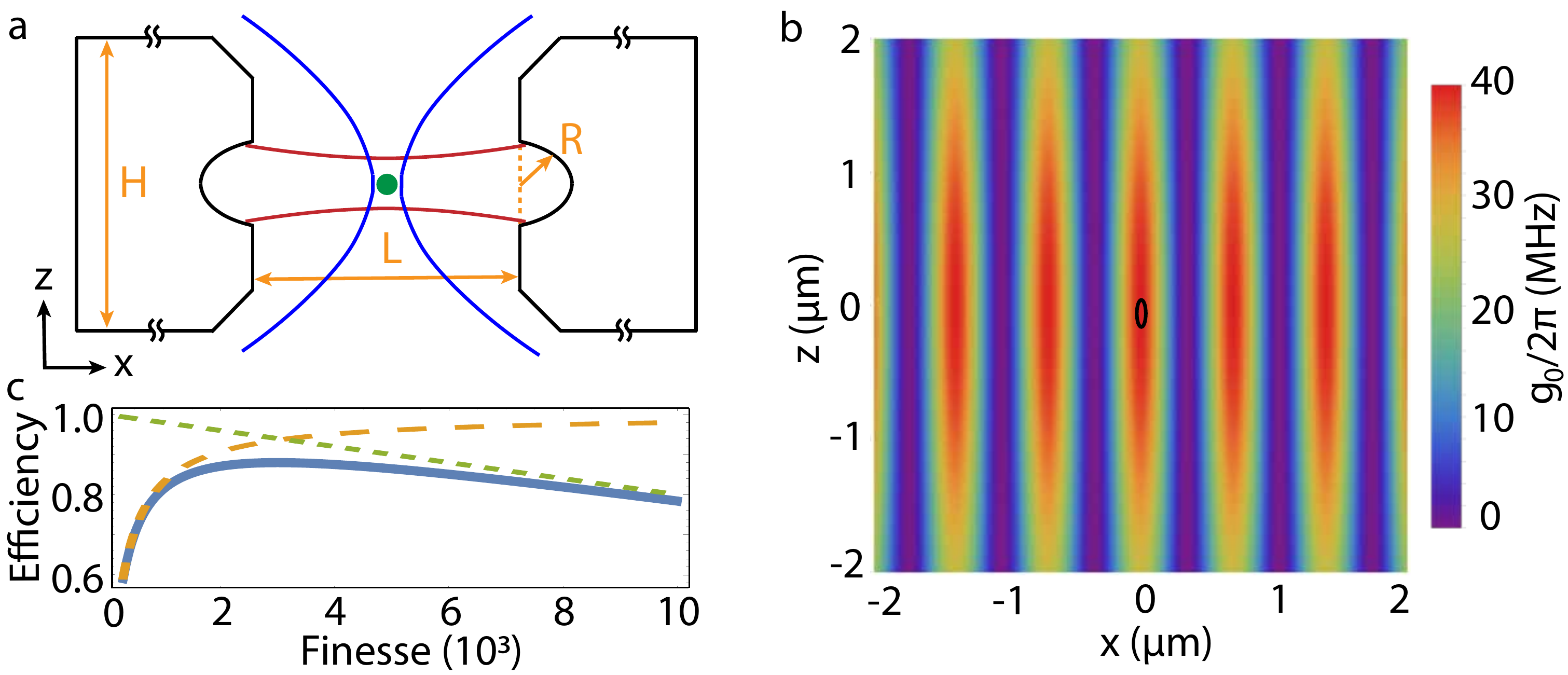}
\caption{Tweezer trapping in a fiber Fabry-P\'erot cavity. (a) The geometry of the fiber FP cavity, and the tweezer trap at the center. (b) A zoom-in of the center $\pm2$ $\mu$m of the cavity mode. The black ellipse shows a liberal estimate of the atomic wavefunction in a tweezer trap as described in the previous section. (c) The efficiencies associated with the fiber cavity vs $\textit{F}_e$ (see text). $P_\text{cavity}$ (orange long dash), $\eta_\text{coll}$ (green short dash), and their product $\eta_\text{ext}$ (blue line).}
\label{fig:Figure3}
\end{figure}

We begin by considering the geometry of fiber FP cavities, as shown in Fig.~\ref{fig:Figure3}a. Typical heights $H$ of the cavity claddings are 125 $\mu$m, so in order to focus a tweezer and image an atom inside with high NA, we assume a cavity length $L=150$ $\mu$m. We choose radii of curvature $R$ of the cavity mirrors to be $R=100$ $\mu$m, which is a typical value for such CO$_2$-laser etching techniques~\cite{Hunger2010}. This geometry at a wavelength of $\lambda=1388.8$ nm gives a mode waist of $\omega_0=4.4$ $\mu$m and a mode volume of $V_m=2.3\times10^{-15}$ m$^3$, or $842\lambda^3$.

We also analyze the mode profile in this cavity in a similar way to the photonic cavity, and a zoom-in of the central $\pm2$ $\mu$m is shown in Fig.~\ref{fig:Figure3}b. The black ellipse shows a liberal estimate of the atomic wavefunction in a tweezer as described in the previous section. The only requirement in this system is that the atom is centered on the antinode as shown in Fig.~\ref{fig:Figure3}b. Noting that the $y$-axis into the page has the same profile as the $z$-direction, this system requires an overall less precise alignment than the photonic cavity.

The coherent coupling rate $g_0$ between the cavity and the atom is given by~\cite{Hunger2010}
\begin{equation}
g_0=\sqrt{\frac{d^2\omega}{2\hbar\epsilon_0V_m}},
\end{equation}
where $d$ is the reduced dipole matrix element, $\omega$ is the angular frequency of the cavity and the atomic transition, $\epsilon_0$ is the permittivity of free space. For this geometry and the value of $d$ discussed above, we obtain a coherent coupling rate of $g_0=2\pi\times39$ MHz. Note that here the cavity supports two degenerate polarizations, so the cavity field can have perfect polarization overlap with the transition dipole. As with the photonic crystal cavity, we design the fiber cavity such that one mirror has lower reflection and allows for coupling photons into and out of the cavity~\cite{Gallego2016,Gallego2018,Welte2018}. Now we choose a finesse of $\mathit{F}_e=2000$, which is well below maximum finesse values of $>10^5$~\cite{Hunger2010}. The cavity linewidth $\kappa_e=\pi~c/(L\mathit{F})$ for these values is $\kappa=2\pi\times500$ MHz, and the free spectral range $\text{FSR}=2\pi~c/(2L)=2\pi\times1.0$ THz. 

We can now estimate the single-atom cooporativity~\cite{Hunger2010}, defined here for consistency with the photonic crystal as $C_0=\frac{4g_0^2}{\kappa\Gamma}$. For our values we arrive at $C_0=39$, for which the probability of emission of a spontaneous photon into the cavity mode is $P_\text{cavity}=C_0/(C_0+1)=0.97$. We assume an intrinsic finesse of $\mathit{F}_i=5\times10^4$~\cite{Hunger2010,Gallego2016}, for which we can again define a collection efficiency $\eta_\text{coll}$ of extracting the photon into the waveguide mode, which is given by $\eta_\text{coll}=1-\textit{F}_e/\textit{F}_i$. For $\textit{F}_e=2000$ this corresponds to $\eta_\text{coll}=0.96$. The efficiency of extracting a photon from the atom into the waveguide mode is given by the product of these $\eta_\text{ext}=P_\text{cavity}\cdot\eta_\text{coll}$, which is 0.94. $\eta_\text{ext}$ is maximized for $\textit{F}_e=2000$, as is shown in Fig.~\ref{fig:Figure3}c.

However, another factor emerges because we require the waveguide mode of the optical fiber to be single-mode (SM). This is necessary since indistinguishable photons are required for Bell state measurements. Representative efficiencies for coupling a photon in a cavity similar to our design into a SM fiber are $\eta_\text{SM}\lesssim0.85$~\cite{Hunger2010}. The photon acquisition efficiency is thus $\eta_\text{PA}=\eta_\text{ext}\cdot\eta_\text{SM}=0.80$.

As with the photonic crystal in the main text, we show a table for the fiber FP cavity of the quantities relevant for a quantum repeater in Table~\ref{tab:Table2}. The Purcell-enhanced linewidth is given by $\Gamma_{1D}=(1+P)\Gamma$, where $P=C_0$. For the fiber FP cavity $\Gamma_{1D}=2\pi\times13$ MHz. For the same $\text{B}_\text{ext}=200$ G and for $\Omega=\Gamma_{1D}$, we get $\Gamma_\text{SC}=2\pi\times390$ kHz, and $\mathcal{F}_\text{RO}>0.99$. 

\begin{table}
\begin{center}
\caption{The values relevant for a quantum repeater with a fiber FP cavity, and the sections of the text in which they are described.}
\label{tab:Table2}
\begin{tabular}{l|c|r}
\textbf{Parameter} & \textbf{Section} & \textbf{Value} \\
\hline
Bandwidth & AIII & 2$\pi\times13$ MHz \\
Wavelength & II & 1.39 $\mu$m (0.35 dB/km) \\
System efficiency & AIII & 0.80 \\
Memory & II, AII & $\leq26$ sec \\
Read-out fidelity & AIII & $>0.99$ \\
\end{tabular}
\end{center}
\end{table} 

\section{Telecom networks with Yb$^+$ ions}
\begin{figure}[t]
\centering
\includegraphics[width=8.6cm]{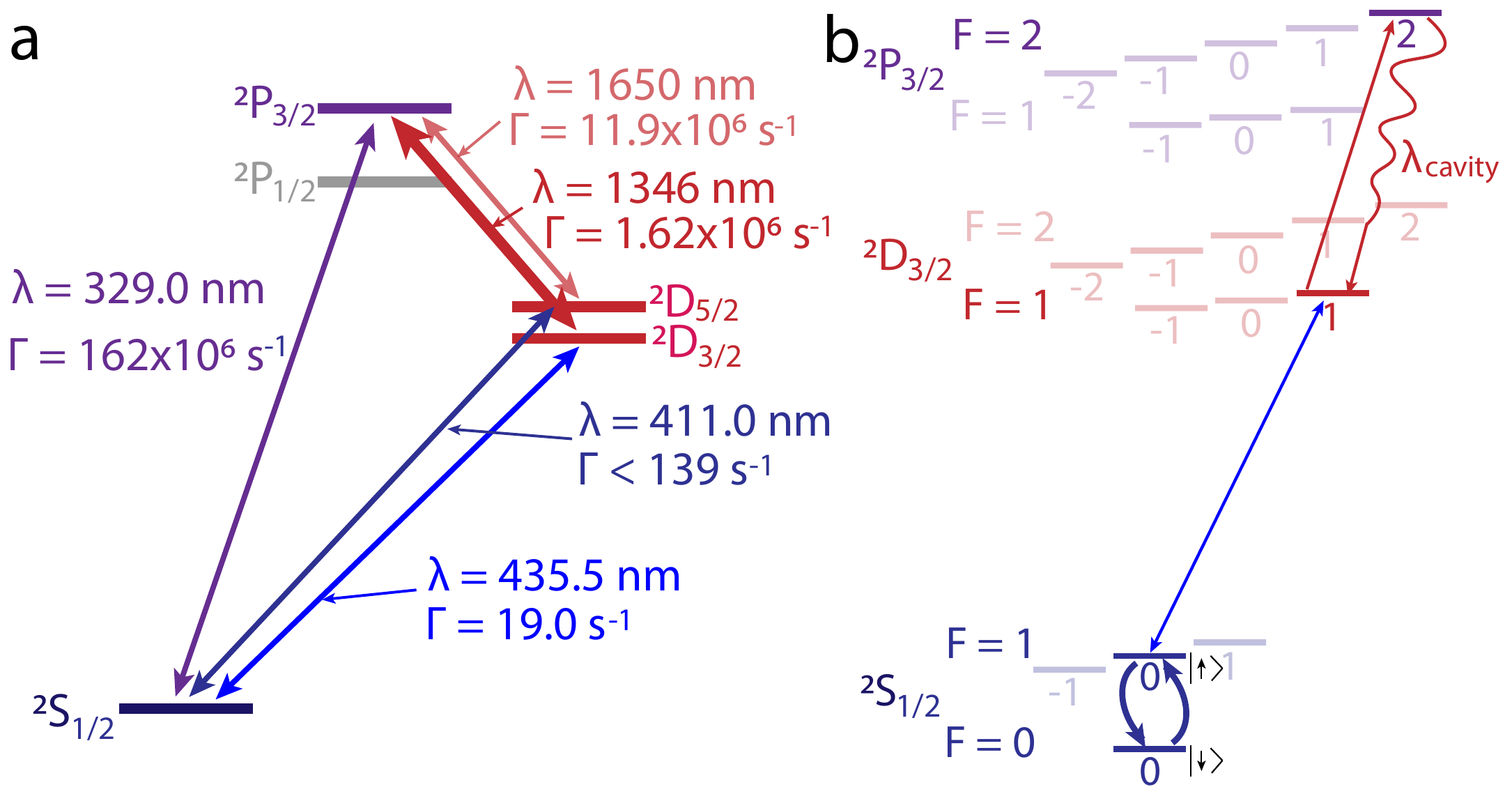}
\caption{Yb$^+$ level structure, and excitation scheme for $^{171}$Yb$^+$. (a) The relavent levels of Yb$^+$, showing their wavelengths and linewidths. The thick red line shows the proposed telecom transition. (b) Three-level scheme by which quantum information in the ground state can be entangled with a telecom photon via the proposad telecom transition resonant with the cavity mode.}
\label{fig:Figure5}
\end{figure}

We now discuss the possibility of using Yb$^+$ ions in fiber FP cavities. Yb$^+$ has strong telecom transitions that remain largely unused in experiments. Yet, they have been studied carefully by theorists because Yb is a platform for parity violation measurements~\cite{Safronova2009,Dzuba2011,Sahoo2011,Porsev2012,Feldker2017}. Similar to the scheme in neutral Yb, the telecom transitions in Yb$^+$ are from metastable states $^2$D$_{3/2}$ and $^2$D$_{5/2}$, which are strongly coupled to a higher-lying state $^2$P$_{3/2}$ via wavelengths of 1346 nm and 1650 nm, respectively. The relevent level structure is shown in Fig.~\ref{fig:Figure5}a. The metastable $^2$D$_{3/2}$ and $^2$D$_{5/2}$ states are very weakly connected to the ground state via `clock'-like transitions, similar to the case of neutral Yb. 

The scheme described in Fig.~\ref{fig:Figure5}b is designed to include the ground-state hyperfine qubit, which has become the workhorse of quantum information processing with trapped ions. One of these qubit states can be mapped to the $^2$D$_{3/2}$ manifold by driving the electric quadrupole (E2) `clock' transition~\cite{Schneider2005}. An excitation from this state to the $2$P$_{3/2}$ allows one of the ground hyperfine qubit states to be entangled with a telecom photon that is strongly coupled to the cavity. The mapping back to the ground state is done by performing a $\pi$-pulse with the `clock' laser.

We choose the $^2$D$_{3/2}$ state rather than $^2$D$_{5/2}$ because it is easier to eliminate other decay pathways from the $^2$P$_{3/2}$ excited state. The hyperfine splitting is 430 MHz for $^2$D$_{3/2}$ but only 64 MHz for $^2$D$_{5/2}$~\cite{Feldker2017}. This latter value is comparable to the Purcell-enhanced linewidth for the decay into the cavity mode. Note that the dipole matrix is 4.2 a$_0$-e for the transition from $^2$P$_{3/2}$ to $^2$D$_{5/2}$ and 1.3 a$_0$-e for the transition to $^2$D$_{3/2}$~\cite{Porsev2012}. The Purcell-enhanced linewidth for the $^2$D$_{3/2}$ state in a fiber FP cavity is well below the hyperfine splitting of 430 MHz. 

Using Equations 2 and 3 above, we can arrive at the exact DME for the specific scheme described in Fig.~\ref{fig:Figure5}b. This gives DME$_{|1,1\rangle\rightarrow|2,2\rangle}=0.41$ a$_{0}$-e. This is smaller than the case for neutral Yb by a factor of $>3$ even though the linewidth $\Gamma$ is similar. This is because $J>0$, and thus there is hyperfine splitting and more states that contribute to the total linewidth.

We believe that the fiber FP cavity considered above and in Fig.~\ref{fig:Figure3} is also suitable for Yb$^+$. Indeed, a Yb$^+$ ion has already been trapped inside a fiber cavity, where the fiber length is $\sim200$ $\mu$m~\cite{Steiner2013}. In that work, the same metastable state $^2$D$_{3/2}$ was used, but it was coupled to an even higher excited state $^3$D$[3/2]_{1/2}$ (not shown), with a transition wavelength of 935 nm. The atomic wavefunction in an ion trap is of similar size to that in an optical tweezer, so the entire discussion above and in Fig.~\ref{fig:Figure4} applies here as well. 

As for the neutral Yb scheme described above, this scheme for Yb$^+$ ion could be used for creating a clock network. This transition is already used for atomic ion frequency standards~\cite{Schneider2005}, and so the development of a commercial technology based on this system may be more straightforward. Further, such a scheme could allow direct telecommunications between trapped ion quantum computers to facilitate the realization of a quantum computing network~\cite{Duan2010}, and it could also be used with multiple qubits at each node of the network, for which gate operations between qubits can be employed for error correction or decoherence free subspaces~\cite{Zwerger2017}.

\end{document}